\documentclass[preprint,floats,aps,epsfig,nofootinbib,amssymb]{revtex4}
\usepackage{mathrsfs}
\usepackage{slashed}
\usepackage{graphicx}
\usepackage{epsfig}
\usepackage{subfigure}
\usepackage{dcolumn}
\usepackage{bm}
\def\gsim{\lower0.5ex\hbox{$\:\buildrel >\over\sim\:$}}
\def\lsim{\lower0.5ex\hbox{$\:\buildrel <\over\sim\:$}}

\begin{document}


\title{The Horava-Lifshitz Type Quantum Field Theory and the Hierarchy Problem}

\author{\bf Wei Chao}

\affiliation{Center for High Energy Physics,
Peking university, Beijing 100871, China \vspace{2.5cm} }

\begin{abstract}
We study the Lifshitz type extension of the standard model (SM) at
the UV, with dynamical critical exponent $z=3$. One loop radiative
corrections to the Higgs mass in such a model are calculated. Our
result shows that, the Hierarchy problem, which has initiated many
excellent extension of the minimal SM, may be weakened in the $z=3$
Lifshitz type quantum field theory.
\end{abstract}
\maketitle

\section{Introduction}
Our current description of the basic interactions in nature, based
on the standard model (SM) of particle physics and general
relativity, is in spectacular agreement with all known experiments.
However, it is almost certainly fundamentally incomplete. The
extreme fine-tuning needed to keep the Higgs mass small compared to
the Planck scale (i.e., the Hierarchy problem) has motivated many
extension of the minimal SM. All of these contain new physics,
beyond the SM, which might be tested at the Large Hadron Collider
(LHC). The most widely explored of these new physics is
supersymmetry (SUSY).

Recently, a new quantum field theory of gravity with ``dynamical
critical exponent" $z$ equal to 3 in the ultra-violet (UV) was introduced in
\cite{horava}, which is called Horava-Lifshitz gravity. 
Although having no complete diffeomorphism invariance of the General Relativity but only the subset(a form of local Galilean invariance),
this theory is power-counting renormalizable
\cite{renormal} in the 3+1 spacetime and can be quantized using
stochastic quantization method \cite{yonwu}. 
The general relativity and Lorentz symmetry in local frame can be recovered in the infrared limit. 
It may provide a ghost free UV complete theory of non-relativistic  gravity around the flat space.
Moreover, it was found that  the evolution of the universe in the Horava-Lifshitz gravity might be singularity free~\cite{Calcagni:2009ar,Kiritsis:2009sh,Brandenberger:2009yt,Cai:2009in}.  
As its characteristic, the theory exhibits
scaling properties which are anisotropic between space and time
\begin{eqnarray}
\mathbf{x} \rightarrow b \mathbf{x} \; , \hspace{1cm} t \rightarrow
b^z t \; .
\end{eqnarray}
Measuring dimensions of operators in the unit of spatial momentum,
one gets $[x]=-1$ and $[t]=-z$. The system doesn't possess the Lorentz
invariance for nontrivial exponent $z$ ($z\neq 1$) but possesses
spatial rotational and translational invariance. The prototype of
such a quantum field theory with $z\neq1$ is the theory of a
Lifshitz scalar in D+1 dimensions, first proposed as a description
of tricritical phenomena. The action of the Lifshitz scalar can be
written as
\begin{eqnarray}
S= \int dt d^D x \left \{ ( \partial_t^{}{\phi})^2 -\lambda (\Delta
\phi)^2 \right \} \; ,
\end{eqnarray}
where $\Delta$ is the spatial Laplacian. This action describes a
free-field fixed point with critical exponent $z=2$. Such fixed
points with anisotropic scale invariance are called Lifshitz points.
Properties of Lifshitz type field theory have been investigated in
\cite{earlerlifshitz}. The construction of gauge theories with
Lifshitz fixed points in D+1 dimensions have been discussed in
\cite{lifshitz,yangmills,yangmills1}. Such a theory flows naturally
to the relativistic value $z=1$ at long distance, and therefore the
Lorentz symmetry shall appear as an emergent symmetry.

In this paper we assume that the SM has a Lorentz non-invariant
UV-completion and realize it at a Lifshitz fixed point with critical
exponent $z=3$. A similar extension of the SM in Lorentz violating
approach is discussed in \cite{lorentzsm,lorentzsm1}. We focus on
the one-loop radiative corrections to the Higgs mass. Our results
show that the Hierarchy problem can be weakened in such a theory.

The outline of the paper is as follows: In section II, we will
construct the SM at the Lifshitz fixed point. In section III, we
will calculate the one-loop radiative corrections to the Higgs mass.
Some conclusions are drawn in Section IV.

\section{The model}
In this section, we will construct the SM at the Lifshitz fixed
point and derive propagators for scalar, vector and spinor fields,
which shall be used for the loop calculations in the next section.
The interactions for the spinor and Higgs fields, with dynamical
critical exponent $z$,  can be written as
\begin{eqnarray}
S_{\rm F}^{}{}&=&\int dt d^3 \vec{x} \bar \psi i \left\{ \gamma^0
D_t^{} - \alpha \gamma^i D_i ( \vec{D} )^{z-1} \right \} \psi \; ,
\\
S_{\rm H}^{}&=& \int dt d^3 \vec x \left \{  ( D_t^{} H)^2 - \beta [
D_i ( \vec{D} )^{z-1} H]^2  - {1 \over 2}{\mu^2} (H^\dagger H)
-\sum_{n=2}^{c} {1 \over 2^n} \lambda_n^{} (H^\dagger H)^{n}
\right\} \; ,
\end{eqnarray}
where $\psi$ represents the ${\rm SU(2)_L^{}}$ doublet or singlet,
$\gamma^i (i=0\sim3)$ is the gamma matrix, $D_\mu^{}(\partial_\mu^{}
+ i g_1^{} \tau^i W_\mu^i + i g_2^{} Y B_\mu^{})$ is the covariant
derivative and $c$ is the biggest integer satisfying the inequality
$c(3-z)\leqslant3+z$. The engineering dimensions of $\psi$ and $H$,
as well as coupling constants $\alpha$ and $\beta$, are given by
\begin{eqnarray}
[\psi]={3\over 2}\; , \hspace{1cm} [H]={3-z\over 2}\; , \hspace{1cm}
[\alpha]=[\beta]=0 \; .
\end{eqnarray}
The system has a free-field fixed point with $z=3$ for any spatial
dimensions. In this case, $c$ in Eq. (4) goes to infinity. The
Yukawa interactions between the Higgs and spinor fields are
\begin{eqnarray}
S_{\rm Y}^{}= \int dt d^3 \vec{x}  \left [ Y_e^{}
\overline{\ell_L^{}} H E_R^{} + Y_u^{} \overline{q_L^{}} \tilde{H}
u_R^{} + Y_d^{} \overline{q_L^{}} H d_R^{} +{ \rm h.c.}\right ] \;.
\end{eqnarray}
Here $[Y_\alpha^{}]=3/2(z-1)$. In Ref. \cite{liffermi}, there is a
novel approach responsible for the origin of fermion mass in a
Lifshitz type extension of the SM including an extra scalar field.
In this paper, we simply assume that Yukawa interactions are
responsible for the origin of fermion masses after the electroweak
symmetry spontaneously broken. For $z=3$, there are two other
renormalizable terms: $\kappa_{gf}^{} \overline{\ell_{\rm L}^C}^g
\varepsilon H {\ell_{\rm L}^{T}}^f \varepsilon H $ and
$\zeta_{gf}^{kl} \overline{f_{\rm L}}^g  H {f_{\rm R}^{f}}
\overline{f_{{\rm L} k}^{}}  H f_{{\rm R} l}^{} $, here $f$ stands
for quarks or leptons and the first term can be used to generate
neutrino tiny Majorana masses without introducing right-handed
Majorana neutrinos.

Now let's construct the gauge field theory with arbitrary dynamical
critical exponent $z$ in 3+1 dimensions. We take $A_0^a$, $A_i^a$ as
the time and spatial components of gauge fields, separately. Our
theory should be invariant under the following gauge transformation:
\begin{eqnarray}
A_\mu^{a} \rightarrow A_\mu^a  + { 1 \over g } \partial_\mu^{}
\varepsilon^a + f^{abc} A_\mu^b \varepsilon^c = A_\mu^a + {1\over g}
D_\mu^{} \varepsilon^a \; .
\end{eqnarray}
Gauge invariant Lagrangian will be constructed from the field
strengths, which are constructed from the commutations of covariant
derivatives as $[D_t^{}, D_i^{}]=ig E_i^{}$ and $[D_i^{}, D_j^{}]=ig
F_{ij}^{}$. The Lagrangian should contain a kinetic term which is
quadratic in first time derivatives. Following the strategy proposed
in Ref. \cite{yangmills}, we obtain the following gauge invariant
interactions
\begin{eqnarray}
S_{\rm YM}^{}= \int dt d^3\vec{x}  \left[ ~{1 \over 2} ~{\rm Tr} (
E_i^{} E_i^{})-{1\over 2 \delta} ~{\rm Tr}\left( \prod_j^{z-1}
D_j^{} F_{ik} \right)^2 ~ \right] \; , \hspace{0.5cm} (z>1) \;
\end{eqnarray}
where ${\rm Tr}$ represents the trace for the gauge generators and
$\delta$ denotes dimensionless coupling constant. The engineering
dimensions of the gauge field components and coupling constants at
the corresponding fixed point $z$ are
\begin{eqnarray}
[A_t^{}]={z+1 \over 2 } \; , \hspace{1cm} [A_i^{}] = {3-z \over 2}\;
, \hspace{1cm} [g_i^{}]={z-1 \over 2} \; .
\end{eqnarray}

The equation of motion for the gauge fields can be obtained from Eq.
(8) by varying $A_0^{}$: $
\partial_t^{} \left( \partial_i^{} A_i^{} \right) - D_i^{} D_i^{}
A_0^{} =0. $  We choose the following natural gauge-fixing
condition:
\begin{eqnarray}
A_0^{}=0 \; , \hspace{1cm} {\rm and } \hspace{1cm} \partial_i^{}
A_i^{} = 0 \; .
\end{eqnarray}
According to the equation of motion, once we adopt the gauge-fixing
condition in Eq. (10) at $t= t_0^{}$, this condition will continue
to hold for all $t$. Then we can derive the propagator for gauge
fields using functional method
\begin{eqnarray}
\langle A_i^a A_j^b  \rangle  \propto {-i  g_{ij}^{} \delta_{ab}^{}
\over k_0^2 - \delta \textbf{k}^6 } \; .
\end{eqnarray}
We can also derive the propagators for spinor and Higgs fields from
Eq. (3) and (4), using the same method:
\begin{eqnarray}
\langle \psi^f_k \overline \psi^g_l \rangle \propto {i \over
\gamma^0 k_0 - \alpha \gamma^i k_i \textbf{k}^2} \delta_{kl}^{}
\delta_{gf}^{} \; , \hspace{1cm } \langle HH^\dagger\rangle \propto
{i \over k_0^2- \beta \textbf{k}^6}\; .
\end{eqnarray}

As can be seen, interactions given in Eqs. (3), (4) and (8) are
incomplete. The full theory should contain all operators with
dimension less than $z+3$, which are not forbidden by symmetries.
Therefore, following terms should be added to the Lagrangian:
$-\alpha_1^{} \overline{\psi} i \gamma^i D_i^{} \psi$, $-\beta_1^{}
|D_i^{} H|^2$, $-1/2 {\rm Tr}(F_{ij}^{})^2$, $(-1/2\delta_1^{}) {\rm
Tr}[D_j^{} F_{ik}^{}]^2 $. That is why we use the symbol
$``\propto"$ instead of $``="$ in Eqs. (11) and (12). However, when
$\Lambda>\Lambda_{\rm IR}^{}$, where $\Lambda_{\rm IR}^{}$ stands
for infrared cut off, below which the Lorentz symmetry is recovered,
these terms will be the subdominant contribution to the propagators.
As a result, we can safely use these propagators to preform one-loop
calculations at $\Lambda(>\Lambda_{\rm IR}^{})$.

A distinctive feature of the Lifshitz type quantum field theory is
that, the dispersion relation becomes $E^2- \sum_i^{z} \alpha_i^{}
\textbf{k}^{2i}=m^2$ \cite{dispersion1}, where $\alpha_i^{}$ are
marginal coupling parameters. Then the speed of light can be written
as
\begin{eqnarray}
c=\left(\sum_n^z  n \alpha_n^{}k^{2n-1}\right)\left(\sum_n^{z}
\alpha_n^{} k^{2n}\right)^{-{1\over2}} \; ,
\end{eqnarray}
where $k=|\textbf{k}|$. In our case, $\alpha_1^{}=1$, $\alpha_2^{}=
\delta_1^{}$ and $\alpha_3^{}=\delta$. One finds that,  if
$z\geqslant2$ the speed of light goes to infinity in the UV. For
$z=3$, the discrepancy of the speed of light at the UV and IR can be
used to explain the time delays in gamma-ray bursts
\cite{yangmills1}.

\section{Hierarchy problem}

For a long time, there were only two solutions to the Hierarchy
Problem: SUSY and technicolor, and SUSY is heavily favored. In
recent years, there are several other new ways to address the
Hierarchy problem, including ADD models \cite{addmodel}, little
Higgs models \cite{littlehiggs}, twin Higgs models \cite{twohiggs},
folded SUSY \cite{foldsusy}, Lee-Wick SM \cite{leewicksm}, and so
on. In this section, we will explore a new solution to the Hierarchy
problem in the Lifshitz type quantum field theory. We assume $z=3$
and then calculate one-loop radiative corrections to the Higgs mass
in such a theory, using the propagators for scalar, spinor and
vector fields presented in Eqs. (11) and (12). Relevant feynman
diagrams are listed in Fig. (1).
\begin{figure}[h]
\includegraphics[scale=0.4]{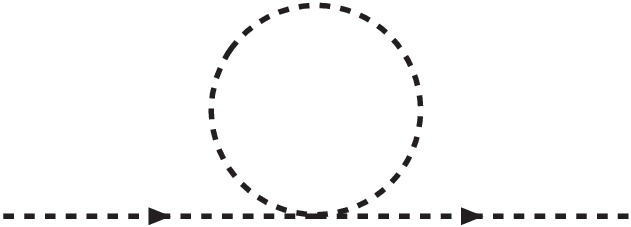}
\includegraphics[scale=0.4]{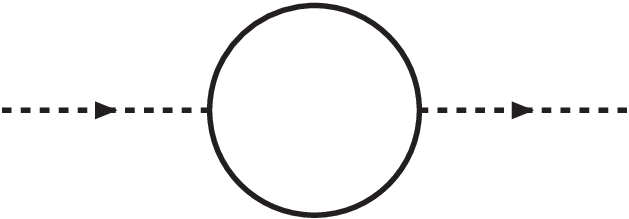}
\includegraphics[scale=0.4]{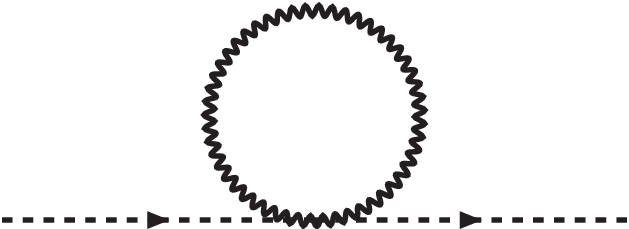}
\\
(a) \hspace{3.6cm} (b) \hspace{3.6cm} (c)
\caption{One-loop radiative corrections to the Higgs mass.}
\end{figure}

Fig. 1 (a) comes from Higgs self interaction. Direct calculation to
this diagram results in
\begin{eqnarray}
\lambda \int {  d^4  k \over (2\pi)^4 } {i \over k^2 - \beta k^6
-\mu^2} = {\lambda \over 24 \pi^2 \sqrt{\beta} } \ln {\Lambda_{\rm
UV}^{} \over \Lambda_{\rm IR}^{} + \mu^2} \; ,
\end{eqnarray}
where $\lambda$ is the four Higgs self-interaction coupling constant
and $\Lambda_{\rm UV}^{}$ is ultraviolet cut off. We may find that
there is only logarithmic divergence instead of quadratic divergence
in Eq. (14). Fig. 1 (b) comes from Yukawa interactions. Direct
calculation to this diagram results in
\begin{eqnarray}
&&\int{ d^4  k \over (2\pi)^4 } (-1) {\rm Tr} \left[Y^\alpha_{kl} {i
\over \gamma^0 k_0 - \alpha \gamma^i k_i  \textbf{k}^2 }\cdot
Y^{\alpha \dagger}_{lk}{ i \over \gamma^0 (k_0^{}-p_0^{})- \alpha
\gamma^i (k_i^{}
-p_i^{} )(\textbf{k-p})^2}\right ] \nonumber \\
&=& {\rm T} \int {d^4k \over (2\pi)^4}
{k_0^{}(k_0^{}-p_0^{})-\alpha^2\textbf{k}\cdot
(\textbf{k}-\textbf{p})\textbf{k}^2(\textbf{k}-\textbf{p})^2
\over (k_0^2-\alpha^2\textbf{k}^6)[(k_0^{}-p_0)^2-
\alpha^2(\textbf{k}-\textbf{p})^6]}\nonumber \\
&\approx&  {{\rm T} \over 24 \pi^2 \alpha} \ln{\left( {\Lambda_{{\rm
UV}^{}}^{} \over \Lambda_{\rm IR}^{}}\right)} \; ,
\end{eqnarray}
where ${\rm T}\equiv -{\rm Tr} [Y_e^{} Y_e^\dagger+ 3Y_u^{}
Y_u^\dagger+ 3Y_d^{} Y_d^\dagger]$. To calculate integral in Eq.
(15), we have used the approximation $p\ll k$, which is good enough
for $p\ll \Lambda_{\rm IR}^{}$. Fig. 1 (c) comes from gauge
interactions of Higgs field, which gives
\begin{eqnarray}
\left ( {3\over 2 }g_2^2 +{1 \over 2 } g_1^2 \right) \textbf{p}^4
\int d^4k {i  \over k_0^2 - \delta \textbf{k}^6}={ \textbf{p}^4
\over 48 \pi^2 \sqrt{\delta}}\left ( 3 g_2^2 + g_1^2 \right)
\ln{\left( {\Lambda_{{\rm UV}^{}}^{} \over \Lambda_{\rm
IR}^{}}\right)} \; ,
\end{eqnarray}
where $\textbf{p}$ is the spatial momentum of Higgs field and
$g_1^{}$, $g_2^{}$ are gauge coupling constants corresponding to
${\rm U(1)}_{\rm Y}^{}$ and ${\rm SU(2)_L^{}}$, respectively.

To sum up, traditional feynman diagrams listed in Fig. 1, that are
quadratic divergent in the SM, become logarithmic divergent in the
$z=3$ Lifshitz type quantum field theory. Actually, in this theory
there are two other feynman diagrams that may contribute to Higgs
mass. We list them in Fig. 2.

\begin{figure}[h]
\includegraphics[scale=0.4]{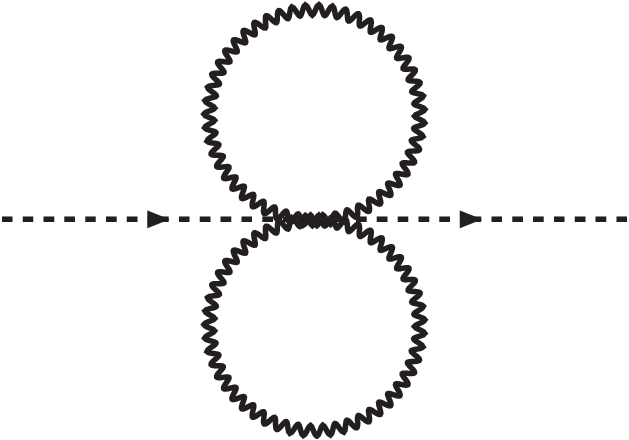}
\includegraphics[scale=0.6]{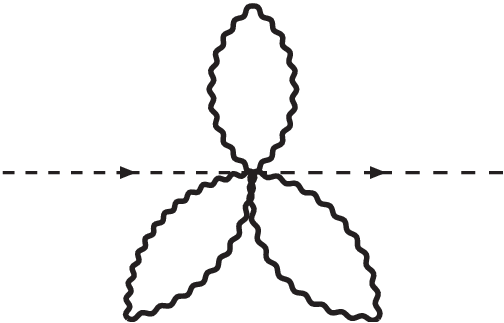}
\\
(a) \hspace{4cm} (b)
\caption{Extra feynman diagrams in the $z=3$ Lifshitz type quantum
field theory, that contribute to the Higgs mass.}
\end{figure}

Direct calculation to Fig. 2 (a) results in
\begin{eqnarray}
{\rm Q} ~\textbf{p}^2 \left[\int d^4k {i \over k_0^2 - \delta
\textbf{k}^6} \right]^2= {\rm Q} ~ \textbf{p}^2   {1 \over 576
\pi^4\delta} \left[ \ln {\left( {\Lambda_{{\rm UV}^{}}^{} \over
\Lambda_{\rm IR}^{}}\right)} \right]^2 \; ,
\end{eqnarray}
where ${\rm Q} \equiv {1 / 4}(6g_2^4 + 3 g_1^2 g_2^2 + g_1^4)$.
Direct calculation to Fig. 2 (b) gives
\begin{eqnarray}
{\rm R} \left[ \int {d^4k \over (2\pi)^4 } {i \over k_0^2 - \delta
\textbf{k}^6} \right]^3={\rm R }  {1 \over 13824 \pi^6 \delta^{3/2}}
\left[ \ln {\left( {\Lambda_{{\rm UV}^{}}^{} \over \Lambda_{\rm
IR}^{}}\right)} \right]^3 \; ,
\end{eqnarray}
where ${\rm R} \equiv {1 / 8 }(10 g^6+ 6 g_1^2 g_2^4 + 3 g_1^4 g_2^2
+ g_1^6 )$. It is clear from Eqs. (17) and (18) that, Fig. 2 (a) and
(b) can not lead to terrible quadratic divergence.

Let's go to investigate the effect on Hierarchy problem from the
interaction of Higgs doublet and four fermions. To guarantee the
gauge invariance, such interaction can be written as
$\zeta_{gf}^{kl} \overline{f_{\rm L}}^g  H {f_{\rm R}^{f}}
\overline{f_{{\rm L} k}^{}}  H f_{{\rm R} l}^{} $. Then the feynman
diagram similar to Fig. 2 (a) with gauge field loop changed with
fermion loop may contribute to the Higgs mass. Actually, fermions
are massless above the electweak scale and such fermionic blob like
feynman diagram does not work at all. We come to the same conclusion
for the interaction, like $\kappa_{gf}^{} \overline{\ell_{\rm
L}^C}^g \varepsilon H {\ell_{\rm L}^{T}}^f \varepsilon H $.

As can be seen in the upper calculations, we only considered the UV
contribution to the Hierarchy problem. Actually, in the
$\Lambda_{\rm IR}^{}$, Lorentz invariance recovers as accidental
symmetry \cite{horava} and Lifshitz type quantum field theory goes
back to the SM. Taking into account the contribution from IR, the
total corrections to the Higgs mass should be
\begin{eqnarray}
\delta m_H^2 \varpropto \mathscr{A} \times \Lambda_{\rm IR}^2 +
\mathscr{B} \times \ln \left( {\Lambda_{\rm UV}^{} \over
\Lambda_{\rm IR}^{} } \right) + \cdots\; ,
\end{eqnarray}
where $\mathscr{A}$ and $\mathscr{B}$ stand for coefficients.  There
are constraints on the scale of Lorentz symmetry violation from HESS
\cite{hess}, MAGIC \cite{magic} and FERMI \cite{Fermia} experiments,
which may be ${ \Lambda_{\rm IR}^{}} \sim 10^{11}~ {\rm GeV}$
\cite{lorentzvo}. Taking this result into Eq. (19), we find that the
Hierarchy problem is still there. But in this case, $\delta m_H^{}$
is proportional to $\Lambda_{\rm IR}^{}$ not to the Planck scale,
such that the Hierarchy problem is weakened.

\section{summary}

In this paper we have considered a Lifshitz type extension of the SM
at the UV with dynamical critical exponent $z=3$. We have written
down the full interactions and derived the propagators for scalar,
spinor and vector fields. Then we have focused on calculating
one-loop radiative corrections to the Higgs mass. Our results show
that,the Hierarchy problem can be weakened in the $z=3$ Lifshitz
type quantum field theory. But still, there are many other problems
for such a theory, which are important and interesting but beyond
the scope of this paper. A detailed study to these topics will be
shown in somewhere else.

\begin{acknowledgments}
The author would like to  thank Professor Z.Z. Xing for constant
encouragement. He is also grateful to Y. Liao and S. Zhou for
helpful discussions. This work was supported in part by the National
Natural Science Foundation of China.
\end{acknowledgments}

\end{document}